 \documentclass[twocolumn,twoside,slac_two]{revtex4}
 \usepackage{graphicx}
 \usepackage{fancyhdr}
\begin{document}

\title{AGILE observations of Supernova Remnants}

\author{A. Giuliani}
\affiliation{INAF-IASF Milano, via E.Bassini 15, 20133 Milano, Italy}
\author{M. Cardillo, M. Tavani}
\affiliation{INAF/IASF-Roma,via Del Fosso del Cavaliere 100, 00133 Roma, Italy}
\affiliation{Dip. di Fisica, Univ.di Roma ``Tor Vergata'', via della Ricerca Scientifica 1, 00133 Roma, Italy}

\author{for the AGILE collaboration}
\affiliation{ }

\begin{abstract}
We will review the crucial AGILE gamma-ray SNR observations
focusing on the evidence of hadronic cosmic-ray acceleration that
has been obtained so far. We discuss data on SNR IC443, W28 and
W44. 
We show that in all cases a consistent model of hadronic acceleration
and interaction with gaseous surroundings can be used to successfully
explain the quite complex morphology and spectral characteristics of the sources.
AGILE, with its crucially important sensitivity near 100 MeV, is equipped to prove the existence of pi-zero emission.
\end{abstract}

\maketitle

\thispagestyle{fancy}

\section{SNR W44}

\begin{figure}[h!]
\begin{center}
\includegraphics[bb=0 0 383 246, scale=0.6]{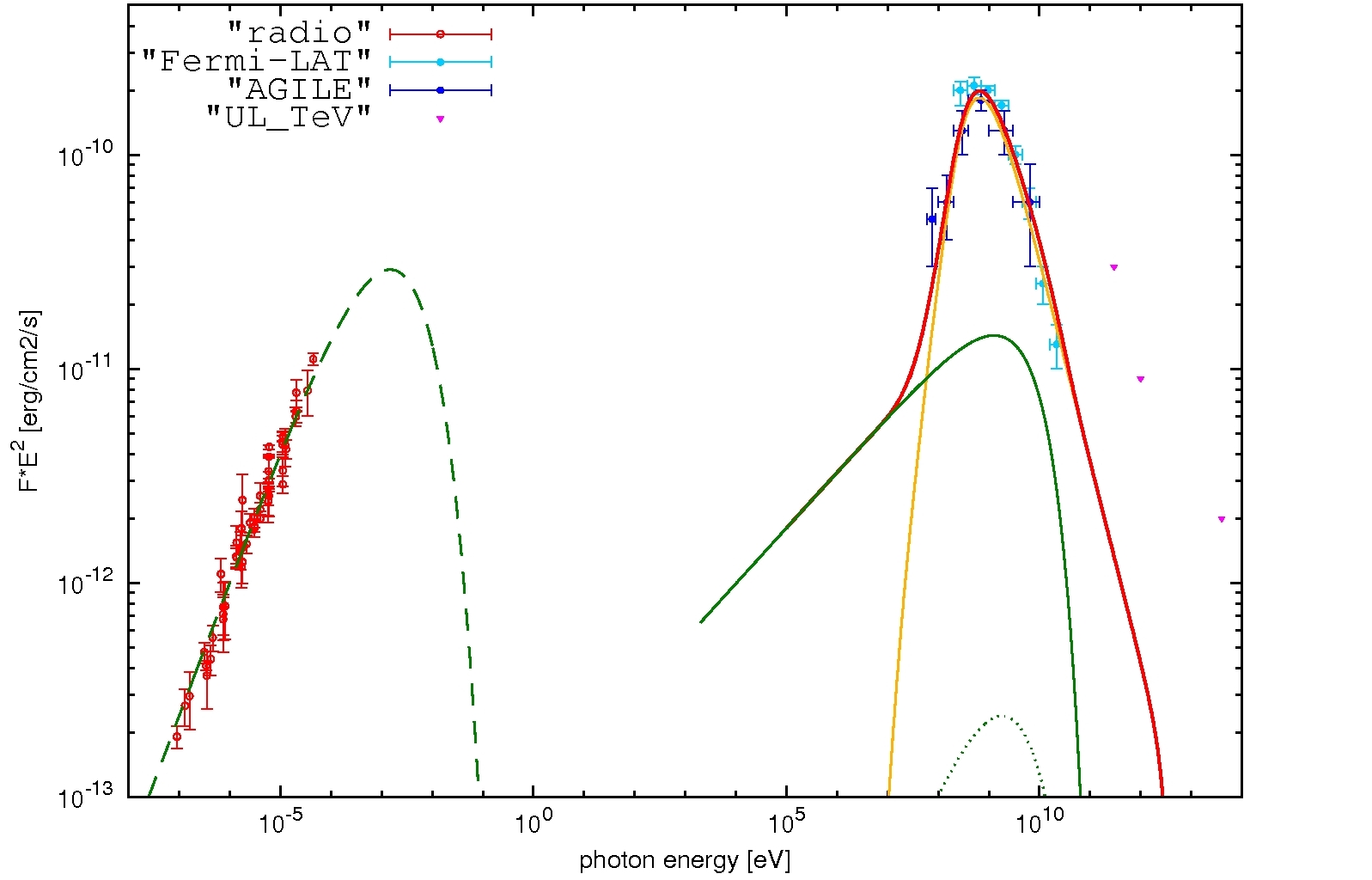}
\end{center}
\caption{Hadronic model, characterized by B
=70 µG and n =100 cm−3, of the broad-band spectrum of
SNR W44 superimposed with the radio (data points in red color)
and gamma-ray data of Fig. 2 (in blue color). TeV upper limits
are also shown. The yellow curve shows the neutral
pion emission from the accelerated proton distribution. The green curves
show the electron contribution by
synchrotron (dashed curve), Bremsstrahlung (solid curve), and IC
(dotted curve) emissions. The red curve shows the total gamma-
ray emission.}
\label{hadronic_model}
\end{figure}

{The middle-aged (~20000 yr) supernova remnants W44 is located at 3.1 kpc from us.}
The AGILE/GRID instrument detected gamma-ray emission from SNR W44 in the energy range 50 MeV - 10 GeV with a significance of 15.8 sigma;
gamma-ray ditribution shows an extended source with a morphology well correlated with the radio shell \cite{giuliani11}.
\begin{figure}[t!]
\begin{center}
\resizebox{5.5cm}{!}{\includegraphics[clip=true]{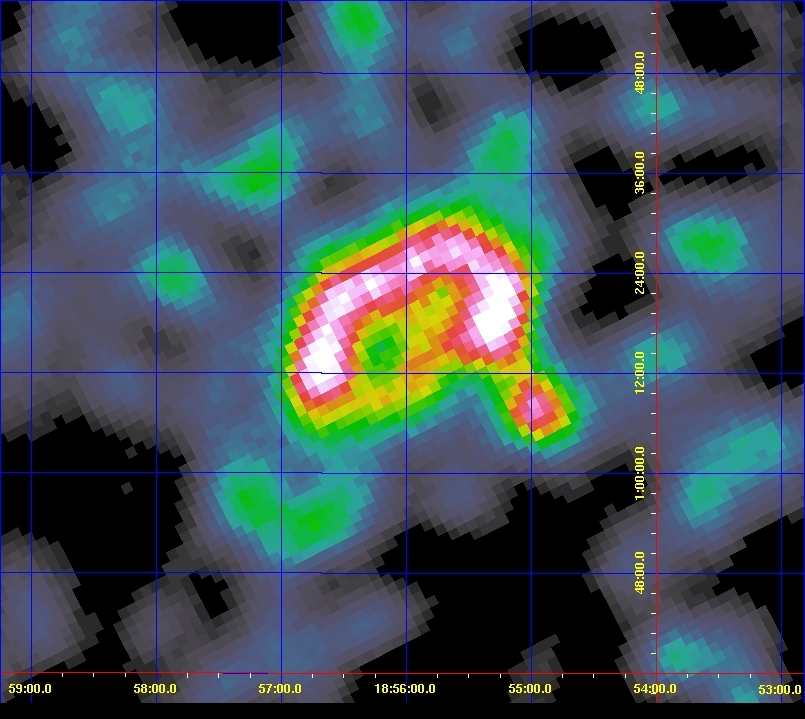}}
\end{center}
\caption{\footnotesize
SNR W44 as seen by AGILE for energies greater than 400 MeV
}
\label{w44}
\end{figure}
The AGILE energy band is complementary with respect to the band 0.2-30 GeV already investigated by the Fermi/LAT instrument for the same
SNR \citep{abdo10}. 
The combination of the AGILE/GRID and Fermi/LAT data allows to obtain a spectrum showing, with unprecedented precision, that the gamma-ray emission from W44 is described by a broad peak around 1 GeV (see figure \ref{hadronic_model}).
With such an accurate spectrum, we can precisely deduce the spectral parameters of the parent particle population (energy of the spectral break, spectral index below and above the break) and dimonstrate that these are not compatible with the electron population seen by radio continuum observations \citep{castelletti07}.\\
An hadronic scenario can instead adequately explain this feature in the gamma-ray spectrum.
In figure \ref{hadronic_model}, AGILE spectral data, together the radio spectral data, are fitted with an hadronic model
characterized by a magnetic field B=70 $\mu$G and a density n =100 cm$^{-3}$. 
A fit of the multi-wavelength data set with leptonic models but no good combination of the parameters exists that can fit W44 spectrum. 

\begin{figure}[h!]
\resizebox{\hsize}{!}{\includegraphics[clip=true]{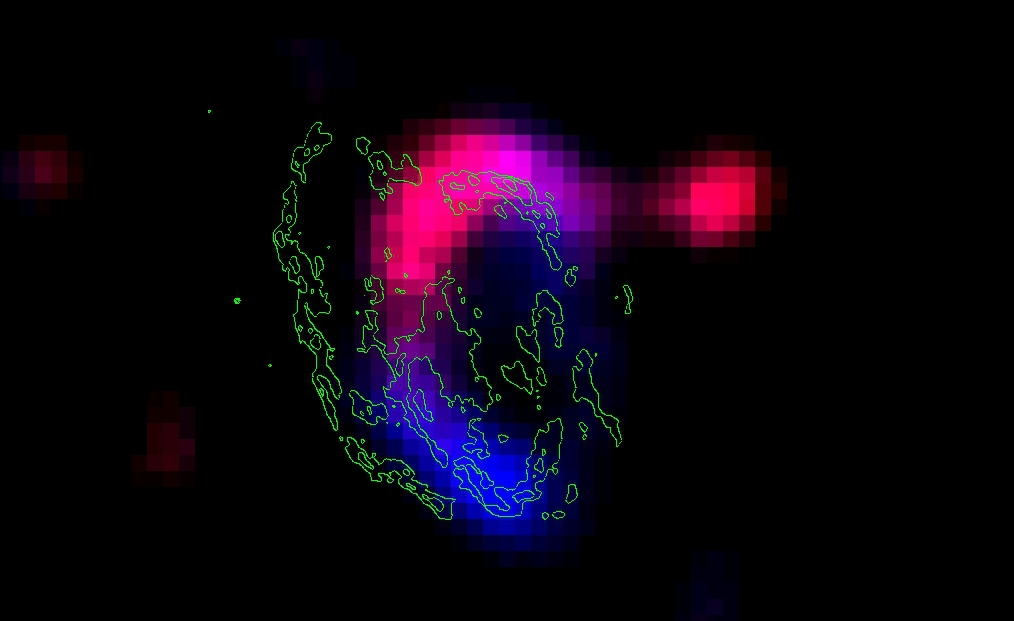}}
\caption{
\footnotesize
AGILE intensity map for the W44 region in the energy range 400MeV-1GeV (Red) and 1-3 GeV (blue). 
Green contours show the radio continuum flux density at 1,4 GHz as observed by the VLA (Castelletti et al, 2007)
}
\label{color}
\end{figure}

AGILE for the first time proved that Galactic Cosmic Rays are accelerated by SNRs.
Moreover, W44 spatial distribution shows that the gamma-ray emission originates in the shell of the SNR.
The large dimensions of the source, combined with the good imaging capabilities of AGILE for E$>$400 MeV, allow us to detect spectral variation
 along the SNR shell.
The energy-resolved spatial distribution shows that the bulk of the emission in the 0.4-1 GeV band is generated in the northern part of
the shell while most of the emission in the 1-3 GeV energy band comes from the southern part of the shell (see figure \ref{color}).
These observations are in agreement with several theoretical models which predict that, at any given time, protons are monochromatically
injected in the ISM with an energy depending on the magnetic field at that time (see, for example, \citealt{gabici09}).
This will allow us to deduce the energy of the protons currently emitted by the different part of the SNR, which is crucial for understanding
the overall acceleration mechanism in SNRs.

\section{SNR W28}

\begin{figure}[b!]
\resizebox{6.5cm}{!}{\includegraphics[clip=true]{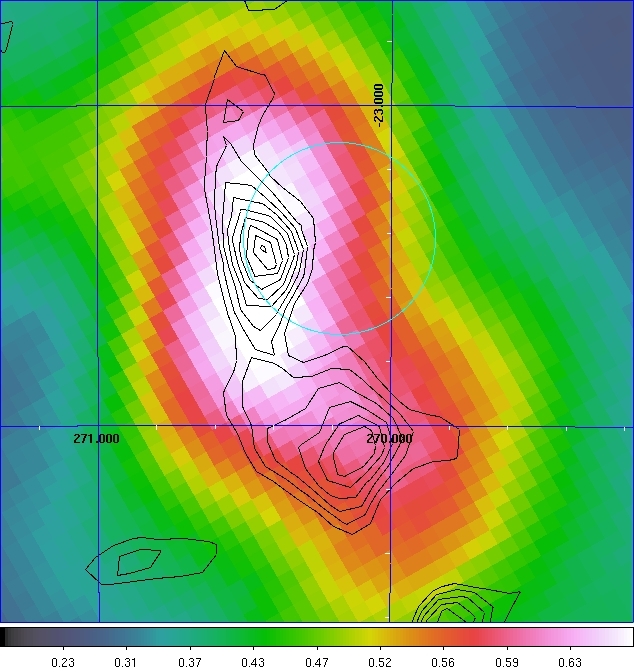}}
\caption{
\footnotesize
AGILE counts map for W28.
The blue circles indicate the location of the supernova remnant W28, the black contours show the CO intensity emission.
}
\label{w28}
\end{figure}

W28 is a mixed morphology SNR with an age of more than 35 000 years located at a distance of about 1.9 kpc. 
A system of massive molecular clouds is associated to the SNR as revealed by the CO (J = 0 $\rightarrow$ 1) observation carried by the
NANTEN telescope \citep{fukui08}.
Two main peaks in the molecular hydrogen distribution can be seen at R.A., dec = 270.4, -23.4 (cloud N) and at R.A., dec = 270.2, -24.1 
(cloud S, see fig. \ref{w28}).
\begin{figure}[b!]
\resizebox{8cm}{!}{\includegraphics[clip=true]{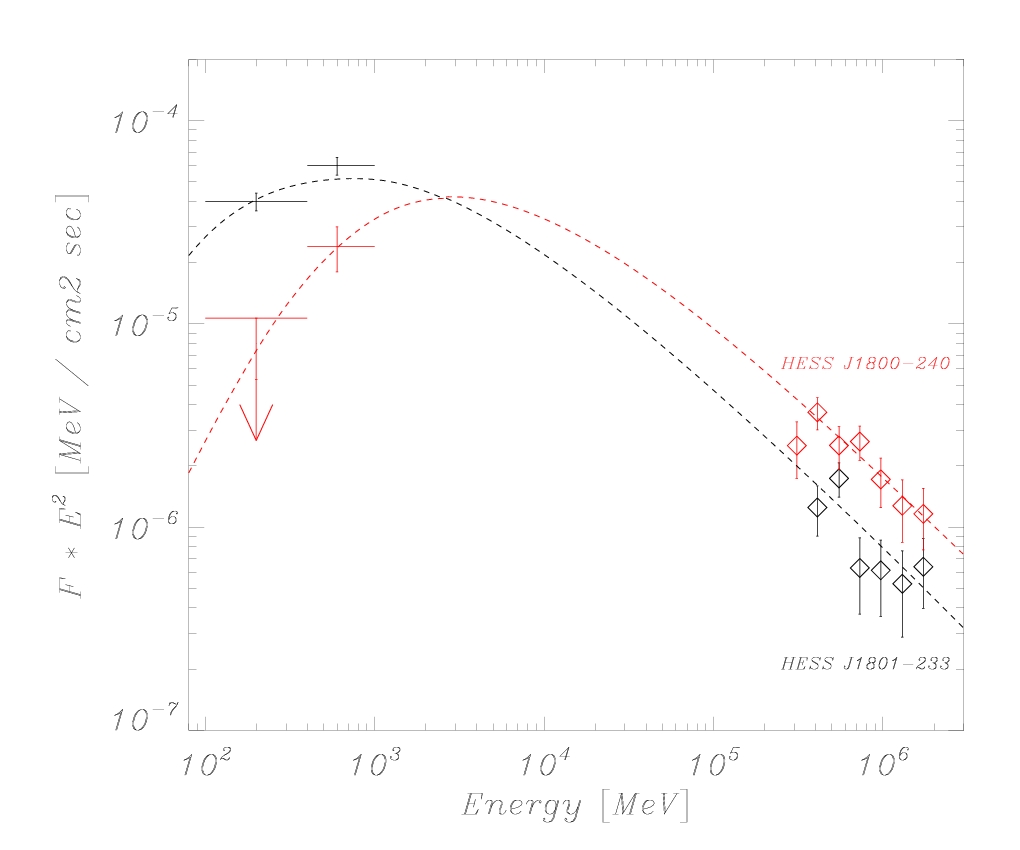}}
\caption{
\footnotesize
Combined AGILE and HESS gamma-ray photon spectra for cloud N
(black) and cloud S (red).
The curves represent the gamma-ray spectra estimated (accordling to the model presented in the text) for the two clouds.
}
\label{w28_spec}
\end{figure}
\begin{figure*}[t!]
\resizebox{15cm}{!}{\includegraphics[clip=true]{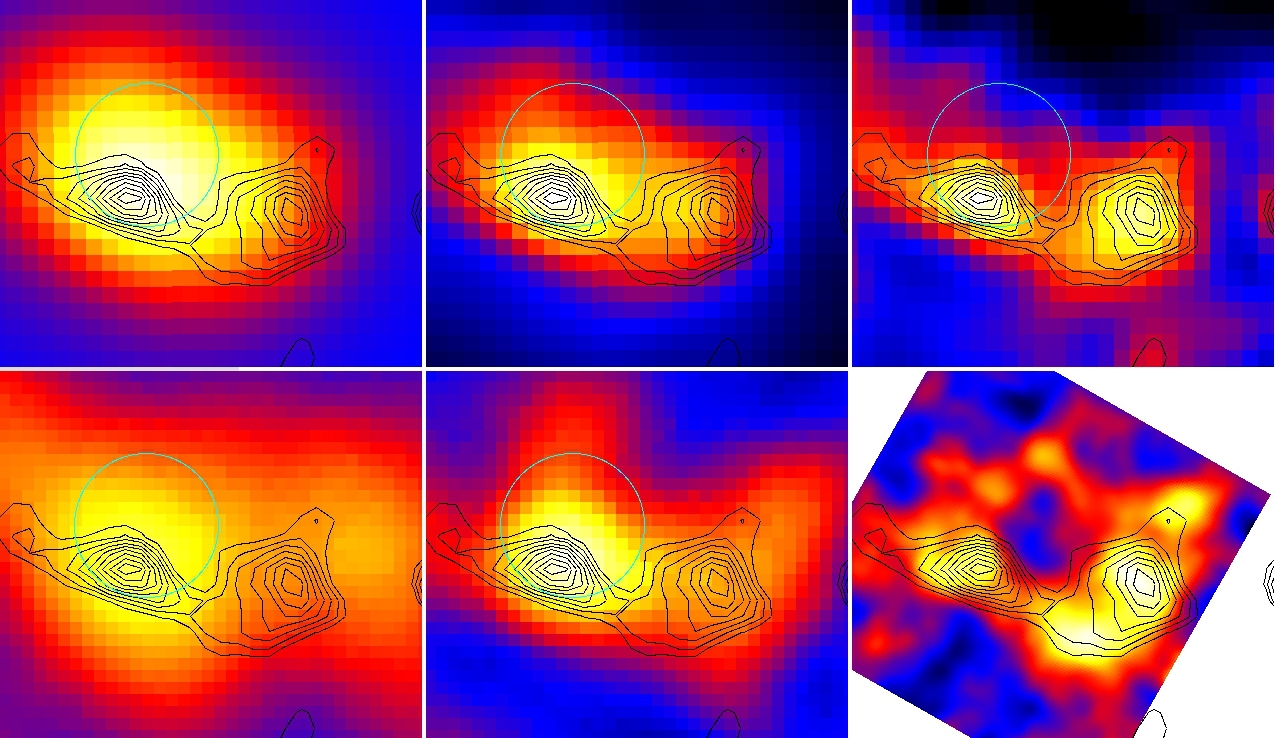}}
\caption{
\footnotesize
\textit{Upper row}: Predicted gamma-ray emission for energies greater than 400 MeV (left column), 3 GeV (center column) and 400 GeV
(right column) for SNR W28.
\textit{Lower row}: Maps produced from the AGILE, Fermi and HESS observations of SNR W28.
}
\label{mevgevtev}
\end{figure*}
Molecular cloud distribution correlates nicely with the gamma ray observations in both the TeV energy band  \citep{aharonian08} and
in the E $>$ 400 MeV energy band observed by AGILE.
However the ratio between the TeV and the multi-MeV emission is significantly different for the cloud N and the cloud S.
In figure \ref{w28_spec} the gamma-ray spectra for the two clouds are shown.
\\The interpretative scenario proposed in \citet{giuliani10} assumes that the N cloud is closer to the CR acceleration site than the S
cloud, considering the energy dependence of the diffusion coefficient.  
If protons diffuse in the interstellar medium with a diffusion coefficient given by $D(E) = D_0 \; E^{0.5}$ the resulting proton energy
spectrum is suppressed below a threshold energy $E_t \sim R^4 t^{-2}$ where R is the distance from the acceleration site and \textit{t}
the age of  the SNR.
Figure \ref{w28_spec} shows the gamma-ray spectra produced by protons (through neutral pion decay) interacting with the cloud N and S
assuming respectively R = 9 and 4 pc.
\\This scenario can explain also the morphology of the gamma-ray emission seen at different energies ranges (see figure \ref{mevgevtev}).
Assuming that CRs are accelerated in a spherical region (indicated by the blue circle), we evaluated their tridimensional distribution,
N(r,E,t), around the SNR, N(r,E,t), as a function of particle energy and SNR age, solving the diffusion equation :
\[
\frac{dN(r,E,t)}{dt} = D(E) \; \nabla^{2} N + \frac{\partial}{\partial E} \left[ b(E) N\right] + Q(E)
\]
where \textit{b(E)} represents the energy losses, and assuming an impulsive injection with a spectrum
$Q(E) \sim E^{-2.2}$.
Black countours in figure \ref{mevgevtev} show the distribution of targets (molecular hydrogen) as derived by the observations of the NANTEN
telescope.
\\The sky maps in the upper row of figure \ref{mevgevtev} refer to the gamma-ray emission for energies greater than 400 MeV, 3 GeV and
400 GeV expected for an age of 40.000 yrs. 
We assumed that gamma-ray emission is produced by p-p collision between accelerated protons and the nuclei of the molecular hydrogen.
In the lower row are shown the maps produced by the AGILE, Fermi and HESS W28 observations.

\section{SNR IC 443}

IC 443 is a SNR lying at a distance of about 1.5 kpc in the Galactic anticenter direction. 
Radio, optical and X-rays emission show a shell structure clearly visible, in correspondence of the interaction between SNR and ISM that
produces a shock.
A system of molecular clouds is also associated to the SNR, and an evidence of the interaction is given by the observation of an high
value of the ratio CO (J=2-1)/(J =1-0) \citep{seta98}.
\\A TeV source has been detected both by MAGIC \citep{albert07} and VERITAS \citep{acciari09}.
Thanks to the good angular resolution of the TeV telescopes it was possible to locate the source in a small error box coincident with
the direction of the most massive cloud.
\\AGILE observed gamma-ray emission in this region obtaining an error box which is not compatible with the MAGIC and VERITAS error boxes
\citep{tavani10}. 
The different position of the source in the TeV and gamma energy ranges implies a difference in the CR spectrum
or in the target distributions. 
A possible interpretation can be given assuming, as in the case of W28, a different distance of the emitting clouds which can lead to a different
spectrum of the accelerated protons seen by the near/far clouds \citep{aharonian06, torres08}.

\begin{table}[h!]
\normalsize
\caption{Middle-aged SNRs seen by~AGILE. Luminosity is given for E $>$ 100 MeV.}
\label{abun}
\begin{center}
\begin{tabular}{|c|c|c|c|}
\hline
SNR             & Age & Distance &  Luminosity  \\
           &   (\textit{years}) & (\textit{kpc}) & (\textit{erg/s})  \\
\hline
W28  	&$ > 35.000 $ 	& $1.9 $		& $3.3\times10^{34} $  \\
\hline
IC 443 	&$ 30.000$ 	& $1.5 $		& $2.4\times 10^{34}$  \\
\hline
W44  	&$ 20.000 $ 	& $2.8 $	 	& $6.5\times10^{34} $  \\
\hline
W51C 	&$ > 20.000 $ 	& $6.0 $		& $7.3\times10^{34} $ \\
\hline

\end{tabular}
\end{center}
\end{table}

\end{document}